\documentclass[aps,10pt,prl,twocolumn,groupedaddress]{revtex4}

\usepackage{graphicx}
\usepackage[caption=false]{subfig}
\usepackage{amsmath}
\usepackage{bbold}
\usepackage{latexsym}

\begin{document}


\title{Particle-hole symmetry without particle-hole symmetry in the quantum Hall effect at $\nu=5/2$. }


\author{P. T. Zucker and D. E. Feldman}
\affiliation{Department of Physics, Brown University, Providence, Rhode Island 02912, USA}


\date{\today}

\begin{abstract}

Numerical results suggest that the quantum Hall effect at $\nu=5/2$ is described by the Pfaffian or anti-Pfaffian state in the absence of disorder and Landau level mixing. Those states are incompatible with the observed transport properties of GaAs heterostructures, where disorder and Landau level mixing are strong. We show that the recent proposal of a PH-Pfaffian topological order by Son is consistent with all experiments. The absence of the particle-hole symmetry at $\nu=5/2$ is not an obstacle to the existence of the PH-Pfaffian order since the order is robust to symmetry breaking.

\end{abstract}



\maketitle


One of the most interesting features of topological insulators and superconductors is their surface behavior. A great variety of gapless and topologically-ordered gapped surface states have been proposed \cite{surface}. Such states are anomalous, that is, they can only exist on the surface of a 3D bulk system and not in a stand-alone film. Finding experimental realizations of exotic surface states proved difficult and most of them have remained theoretical proposals. Thus, it came as a surprise when Son \cite{son} argued that one such exotic state  \cite{hall-top}, made of Dirac composite fermions with the particle-hole symmetry (PHS), has long been observed experimentally in a two-dimensional system: the electron gas in the quantum Hall effect (QHE) with the filling factor $\nu=1/2$.

At first sight, Son's idea violates the fermion doubling theorem \cite{doubling}. However, the theorem does not apply to interacting systems such as the one Son considered. Besides, in contrast to the conditions of the doubling theorem, the action of PHS is nonlocal in QHE since it involves filling a  Landau level. Interestingly, the picture of composite Dirac fermions sheds light on the geometrical resonance experiments \cite{gr} which the classic theory \cite{HLR} of the $1/2$ state could not explain. In this paper we show that a closely related idea \cite{son} provides a natural explanation of the observed phenomenology on the enigmatic QHE plateau at $\nu=5/2$ in GaAs.

Cooper pairing of Dirac composite fermions in the $s$ channel results in a fractional QHE state dubbed PH-Pfaffian \cite{son,ts}, where ``PH" stands for ``particle-hole". We argue that the PH-Pfaffian topological order is present on the QHE plateau at $\nu=5/2$. This might seem unlikely because the particle-hole symmetry is violated by the Landau level mixing (LLM) in the observed states of the second Landau level in GaAs. Besides, numerics 
\cite{morf98,dimov08,feiguin2009,biddle,wojs2010,new-num} 
supports the Pfaffian \cite{2} and anti-Pfaffian \cite{3,4} states at $\nu=5/2$ even in the presence of PHS. At the same time, the existing numerical work, which always neglects strong disorder and typically neglects strong LLM, is not yet in the position to explain the physics of the $5/2$ state, as is evidenced by a large discrepancy between numerical and experimental energy gaps \cite{dAmbrumenil2011}. 
Note that a recent attempt to incorporate LLM \cite{new-num} into simulations led to the manifestly wrong conclusion that the $5/2$ state does not exist at realistic LLM. 
Indeed, as discussed in Ref. \cite{new-num}, the existing perturbative methods are justified only at weak LLM.
There is also no conflict between the PH-Pfaffian order and the absence of PHS at $\nu=5/2$ since topological orders in the fractional QHE are not protected by symmetry. On the other hand, it turns out that the PH-Pfaffian order naturally explains the experimental data that could not be explained \cite{baer,yf14} by the Pfaffian and anti-Pfaffian hypotheses. Thus, the support for the Pfaffian and anti-Pfaffian states from numerical studies of simplified Hamiltonians with PHS opens an intriguing possibility of ``symmetry from no symmetry": the particle-hole Pfaffian order is stabilized by LLM and impurities which break PHS.

Below we review the properties of the PH-Pfaffian state and compare them with experiment.  We also propose new experimental probes. Besides QHE in GaAs, our motivation comes from the possibility of new non-Abelian states \cite{jain16} in ZnO \cite{zno} and multilayer graphene \cite{graphene1,graphene2}.

Most experiments probe edge physics. The edge theory can be constructed from the bulk wave function according to the bulk-edge correspondence \cite{wen}. We will take the opposite route and deduce the PH-Pfaffian ground state from the edge Lagrangian. The latter is fixed by symmetry considerations. We do not expect the ground state to be invariant under PHS. Nevertheless, its topological order must be compatible with such a symmetry because our system is in the same phase as the particle-hole symmetric superfluid built by Cooper pairing of Son's composite fermions. In particular, the electric and thermal Hall conductances are determined by the topological order and hence must be invariant to the particle-hole conjugation.

As usual, we ignore the filled states of the first Landau level. The particle-hole conjugation then demands reversing the directions of all edge modes and adding another integer edge mode with the Hall conductance $e^2/h$ and the thermal conductance $\pi^2 k_B^2 T/3h$. It thus follows from symmetry that the electric and thermal conductances of the PH-system must be one half of the above expressions: $G=e^2/2h$ and $\kappa=\pi^2 k_B^2 T/6h$. This corresponds to an edge theory with a downstream charged Bose-mode whose thermal conductance is $2\kappa$ and an upstream Majorana fermion whose thermal conductance is one half that of a Bose mode. The Lagrangian density

\begin{equation}
\label{1}
\mathcal{L}=-\frac{2}{4\pi}[\partial_t\phi\partial_x\phi+v_c(\partial_x\phi)^2] +i\psi(\partial_t-v_n\partial_x)\psi,
\end{equation}
where the chiral Bose-field $\phi$ propagates with the velocity $v_c$ and determines the charge density on the edge according to $\rho(x)=e\partial_x\phi/2\pi$. The neutral chiral Majorana fermion $\psi=\psi^\dagger$ travels in the direction opposite to that of the Bose-mode. The action (\ref{1}) is very similar to the edge theory of the Pfaffian state \cite{7} and differs only by the propagation direction of the neutral mode.

Many ground-state wave functions correspond to the same low-energy edge theory (\ref{1}). They depend on details of impurities and LLM in a particular sample. We use the generalized Moore-Read prescription \cite{hall-rev} to write an example of a wave function with the PH-Pfaffian topological order. A topological order in the first Landau level is encoded by a simpler wave function than the same order in the second level. To facilitate a comparison with the literature on possible $5/2$ states \cite{yf13} we write a wave function for electrons in the first Landau level:

\begin{equation}
\label{2}
\Psi(\{z_i\})=\int \{d^2\xi_i\}\langle\{z_i\}|\{\xi_i\}\rangle\Phi(\{\xi_i\}),
\end{equation}
where $z_k=x_k+iy_k$ and $\xi_k$ are complex coordinates, $\langle\{z_i\}|\{\xi_i\}\rangle=\Pi_i\exp[-(|\xi_i|^2-2\bar{\xi_i}z_i+|z_i|^2)/(4l_B^2)]$ is the coherent state kernel that projects the wave function into the lowest Landau level, $l_B$ is the magnetic length and the bar denotes complex conjugation. 
The factor $\Phi$ is the correlation function \cite{hall-rev} of the electron operators $\Psi_e=\psi(x)\exp(2i\phi(x))$ in the conformal field theory (\ref{1}):

\begin{equation}
\label{3}
\Phi(\{\xi_i\})={\rm Pf}\left\{\frac{1}{\bar{\xi_i}-\bar{\xi_j}}\right\}\Pi_{i<j}(\xi_i-\xi_j)^2.
\end{equation}
Our choice of the electron operator determines the shift \cite{shift1,shift2} $S=\chi(h_z-h_{\bar z})$, where $\chi$ is the Euler characterstics of the surface that confines electrons, and $h_z$ and $h_{\bar z}$ are the scaling dimensions of the holomorphic and antiholomorphic parts of the electron operator $\Psi_e$. Thus, $S=1$ on a sphere in agreement with Son \cite{son}.

Quasiparticles braid trivially with electrons and are created by the same operators as in the Pfaffian theory \cite{7}. There are six superselection sectors: vaccum $1$, neutral fermion $\psi$, two charge-$e/2$ excitations $\exp(i\phi)$ and $\psi\exp(i\phi)$, and two non-Abelian quasiparticles $\sigma\exp(i\phi/2)$ and $\sigma\exp(3i\phi/2)$ with charges $e/4$ and $3e/4$, where the operator $\sigma$ twists the boundary conditions for the Majorana fermion. The fusion rules are $\psi\times\psi=1$, $\psi\times\sigma=\sigma$ and $\sigma\times\sigma=1+\psi$. The braiding rules are different from the Pfaffian state. We will need the statistical phase, picked up by an $e/4$-particle after it encircles an excitation with the electric charge $ne/4$ and the topological charge $\alpha=1,\sigma$ or $\psi$. The phase depends on the fusion channel $\beta$ of the topological charges $\alpha$ and $\sigma$ and equals

\begin{equation}
\label{4}
\phi=\frac{n\pi}{4}+\phi'_{\alpha\beta},
\end{equation}
where $\phi'_{1\sigma}=0$, $\phi'_{\epsilon\sigma}=\pi$, $\phi'_{\sigma 1}=\pi/4$ and $\phi'_{\sigma\epsilon}=-3\pi/4$.

The particle-hole conjugation changes the signs \cite{3,footnote-phase} of all statistical phases ${\rm mod}~2\pi$. One easily verifies that the above fusion and braiding rules are compatible with PHS. For example, one can use the invariance of the braiding phase under a simultaneous change of the signs of all electric charges. Next, one observes that all phases change their signs ${\rm mod}~2\pi$ after the excitations $\exp(i\phi)$ and $\psi\exp(i\phi)$ transform into each other while all other superselection sectors remain unchanged (remember that $\pi=-\pi~{\rm mod}~2\pi$).

We now turn to comparison with experiment. The experiment \cite{36} revealed an upstream neutral mode on the $5/2$ edge. Upstream modes can be either topologically protected, as at $\nu=2/3$, or emerge from edge reconstruction, as at $\nu=1/3$. The low-temperature propagation length is finite for neutral modes on a reconstructed edge \cite{footnote-edge}. The propagation length diverges at $T\rightarrow 0$ for topologically protected 
modes \cite{footnote-edge, KF}.
The observed propagation length of the neutral mode at $\nu=5/2$ is comparable with the propagation lengths at $\nu=2/3$ and $3/5$ and much longer \cite{proliferation} than at $\nu=1/3$ and $2/5$ in similar samples. Thus, the $5/2$ upstream mode is topologically protected. This agrees with the Lagrangian (\ref{1}). The anti-Pfaffian state also has topologically protected upstream modes. The Pfaffian order appears incompatible with the experiment. Note that a recent numerical study \cite{new-num} supports a protected upstream mode at strong LLM.

Next, consider experiments on quasiparticle tunneling through narrow constrictions \cite{baer,20,21}. Tunneling is dominated by the lowest-charge quasiparticles $\sigma\exp(i\phi/2)$. Theory \cite{wen,yf13} predicts the power dependence of the zero-bias conductance on the temperature: $G\sim T^{2g-2}$ with a universal exponent $g$. The exponent $g=1/4$ is the same in the PH-Pfaffian and Pfaffian states \cite{yf13}. The anti-Pfaffian order corresponds \cite{yf13} to $g=1/2$. Experimental results for $g$ exceed \cite{baer,yf13,rodaro,papa04} the theoretical values at all fractional filling factors \cite{footnote}. This is explained by a combination of three mechanisms that suppress low-temperature tunneling: Coulomb repulsion across the constriction \cite{papa04,yf13}, edge reconstruction \cite{ros-hal,yang03} and dissipation \cite{dis}. Hence, experiments can only give an upper bound on $g$. 
At $\nu=5/2$ that bound \cite{baer,20,21} is $0.4$. Thus, the PH-Pfaffian state is compatible with the tunneling data and the anti-Pfaffian state is not. We observe that the transport data exclude both the Pfaffian and anti-Pfaffian states. The PH-Pfaffian topological order is consistent with the existing experiments.

\begin{figure}[b]
\centering
\includegraphics[width=2in]{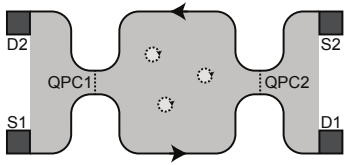}
\caption{Aharonov-Bohm interferometer. Quasiparticles move along the edges and tunnel between the edges at the quantum point contacts QPC1 and QPC2. Several quasiparticles are localized inside the device.}
\label{Fig1}
\end{figure} 

One interpretation of the Fabry-P\'erot interference experiment \cite{willett10} at $\nu=5/2$ is based on the even-odd topological effect \cite{11,12}, predicted for the Pfaffian and anti-Pfaffian states. The same even-odd effect occurs in the PH-Pfaffian state. Indeed, the magnetic flux dependence of the current through a Fabry-P\'erot device, Fig. 1, comes from the interference of the quasiparticle paths through the two constrictions. Such interference is present whenever an even number of $\sigma\exp(i\phi/2)$-quasipaticles is localized inside the device. If the number is odd, the tunneling quasiparticle can fuse with the topological charge $\alpha$ of the interferometer in two ways. In both fusion channels $\beta$, the phase difference $\phi(\beta)$ between the two trajectories is given by the braiding phase (\ref{4}). Since the phases (\ref{4}) differ by $\pi$ in the two fusion channels, there is no interference and the even-odd effect is observed.

In common with the Pfaffian and anti-Pfaffian states, we expect the PH-Pfaffian topological order to occur in a spin-polarized electron liquid. The existing data on the polarization of the $5/2$ liquid are controversial. A recent observation of the $5/2$ plateau in ZnO strengthens the case for nonzero polarization \cite{zno}.

What new experiments could probe the PH-Pfaffian order? First, the thermal Hall conductance $\kappa=\pi^2 k_B^2 T/6h$ in the PH-Pfaffian state differs from all other proposed topological orders. A more striking manifestation of the PH-Pfaffian state comes from Mach-Zehnder interferometry \cite{newmz1,newmz2,mz1,mz2,mz3,mz4,newmz3,mz5} which we address below. We discover two unique signatures of the PH-Pfaffian order: the current through the interferometer does not depend on the magnetic field; the Fano factor for the current noise diverges at some values of the magnetic field. Neither feature is known to occur in any other QHE state at any filling factor.

\begin{figure}[b]
\centering
\includegraphics[width=2.5in]{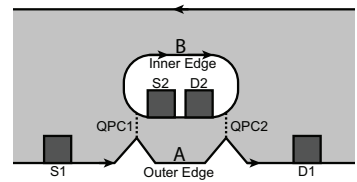}
\caption{Mach-Zehnder interferometer.}
\label{Fig2}
\end{figure} 

Quasiparticles can tunnel between the inner and outer edges of the Mach-Zehnder interferometer, Fig. 2, at quantum point contacts QPC1 and QPC2. After a tunneling event, a quasiparticle/hole travels along the inner edge and is absorbed by drain D2. The drain is inside the interferometer. As a result, the total topological charge, accumulated in the drain, affects the probability of the next tunneling event. Indeed, the phase difference between the two possible tunneling processes through QPC1 and QPC2 depends on the statistical phase $\phi_s$, accumulated when a quasiparticle encircles the drain. The tunneling probability also depends on the Aharonov-Bohm phase $\phi_{\rm AB}$, accumulated on the loop QPC1-A-QPC2-B-QPC1. Since tunneling is dominated by $e/4$-particles, the phase $\phi_{\rm AB}=\pi\Phi/2\Phi_0$, where $\Phi$ is the magnetic flux through the loop QPC1-A-QPC2-B-QPC1 and $\Phi_0=hc/e$ is the flux quantum.

To compute the tunneling probabilities we use the Hamiltonian of the interferometer

\begin{equation}
\label{5}
\hat H=\hat H_{\rm inner}+\hat H_{\rm outer}+\hat T_1+\hat T_2,
\end{equation}
where $\hat H_{\rm inner}$ and $\hat H_{\rm outer}$ are the Hamiltonians of the two edges and the operators $\hat T_{1,2}$  describe quasiparticle tunneling at QPC1 and QPC2. The four terms on the right hand side of Eq. (\ref{5}) depend on the gauge choice. We select such a gauge for the electromagnetic field that all information about the Aharonov-Bohm phase $\phi_{\rm AB}$ is contained in the tunneling operator $\hat T_2$. All information about the statistical phase $\phi_s$ is also absorbed into $\hat T_2$. Thus, we select the tunneling operators in the form

\begin{eqnarray}
\label{6}
\hat T_1=\Gamma_1\Psi^q_{\rm outer}({\rm QPC1})\Psi^{q\dagger}_{\rm inner}({\rm QPC1})+{\rm h.c.}, & & \\
\label{7}
\hat T_2=\Gamma_2\exp(i\phi_{\rm AB}+i\phi_s)\Psi^q_{\rm outer}({\rm QPC2})\Psi^{q\dagger}_{\rm inner}({\rm QPC2})+{\rm h.c.},
\end{eqnarray}
where h.c. stays for the Hermitian conjugate and the operators $\Psi^q_{\rm edge}$ destroy quasiparticles of charge $e/4$ on the inner/outer edge at the locations of the quantum point contacts.

We will restrict our discussion to the zero-temperature limit, where charge only travels from the higher to lower potential. We thus assume that quasiparticles can only tunnel from the outer edge to the inner edge. The tunneling probability $P(ne/4,\alpha,\beta)=u_{\alpha\beta}P^\beta(ne/4,\alpha)$, where $ne/4$ and $\alpha$ are the electric and topological charges of drain D2 before tunneling, $\beta$ is the fusion channel of $\alpha$ with the tunneling quasiparticle,
$u_{\alpha\beta}$ is the probability of the fusion outcome $\beta$, and $P^\beta(ne/4,\alpha)$ is the tunneling probability in the fusion channel $\beta$. $u_{\sigma 1}=u_{\sigma\epsilon}=1/2$. The fusion probabilities of all other possible processes equal 1. The probability $P^\beta$ can be found from the second order perturbation theory in $\hat T_{1,2}$:

\begin{equation}
\label{8}
P^\beta(ne/4,\alpha)=r\{|\Gamma_1|^2+|\Gamma_2|^2+2u|\Gamma_1\Gamma_2|\cos(\phi_{\rm AB}+\phi_s+\gamma)\},
\end{equation}
where $\gamma={\rm arg}[\Gamma_2/\Gamma_1]$, and $r$ and $u$  come from the voltage-dependent correlation functions of the quasiparticle operators in the edge theory with the unperturbed Hamiltonian $\hat H_{\rm inner}+\hat H_{\rm outer}$. At low voltages $eV\ll \hbar v_c/L,~\hbar v_n/L$, where $L$ is the interferometer size, the factor $u\approx 1$ since the distance between QPC1 and QPC2 can be neglected in the calculation of the correlation functions. It is convenient to label the possible values of $P^\beta(ne/4,\alpha)$ as $p_0$, $p_{\pi/2}$, $p_{\pi}$, and $p_{-\pi/2}$ depending on $\phi_s=0$, $\pi/2$, $\pi$, or $-\pi/2$. Fig. 3 shows all the possible transitions between the 6 superselection sectors of the drain.

\begin{figure}[b]
\centering
\includegraphics[width=2.5in]{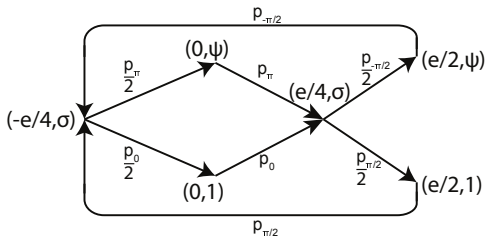}
\caption{The nodes show the superselection sectors of drain D2. The arrows show all the possible transitions between the sectors. The transition probabilities are written next to the arrows.}
\label{Fig3}
\end{figure} 

We are ready to compute the current $I$ between S1 and D2. Be definition, $I=Q/t$, where $Q=Ne$ is the charge, transmitted through the interferometer during a long time interval $t$. At large $N,$ the interval $t\approx N\bar t$, where $\bar t$ is the average time of four consecutive tunneling events. Since each tunneling event transfers charge $e/4$, one finds $I=e/\bar t$.

The system follows the arrows in Fig. 3. Hence, the drain always returns to the initial superselection sector $(-e/4,\sigma)$ after four quasiparticles tunnel. This can happen in one of the four ways: $(-e/4,\sigma)\rightarrow(0,1)\rightarrow(e/4,\sigma)\rightarrow(e/2,1)\rightarrow(-e/4,\sigma),$ $(-e/4,\sigma)\rightarrow(0,\psi)\rightarrow(e/4,\sigma)\rightarrow(e/2,1)\rightarrow(-e/4,\sigma),$ $(-e/4,\sigma)\rightarrow(0,1)\rightarrow(e/4,\sigma)\rightarrow(e/2,\psi)\rightarrow(-e/4,\sigma),$ $(-e/4,\sigma)\rightarrow(0,\psi)\rightarrow(e/4,\sigma)\rightarrow(e/2,\psi)\rightarrow(-e/4,\sigma).$ The average time $\bar t=\sum q_i \bar t_i$, where 
$q_i$ are the probabilities and $\bar t_i$ are the average times for the above four trajectories. For example, the probability of the first trajectory is 
$p_0/[p_0+p_{\pi}]\times p_{\pi/2}/[p_{\pi/2}+p_{-\pi/2}]$. The average travel time along that trajectory is $2/(p_0+p_\pi)+1/p_0+2/(p_{\pi/2}+p_{-\pi/2})+1/p_{\pi/2}$. One finds the electric current

\begin{equation}
\label{9}
I=\frac{e}{16}(p_0+p_{\pi/2}+p_\pi+p_{-\pi/2})=\frac{er}{4}[|\Gamma_1|^2+|\Gamma_2|^2].
\end{equation}
The current does not depend on the magnetic flux through the interferometer.

We next compute the low-frequency current noise $S=\int_{-\infty}^\infty dt[\langle \hat I(0)\hat I(t)+\hat I(t) \hat I(0)\rangle- 2\langle \hat I(0)\rangle^2]$. It can be conveniently represented as 
$S=2\langle\delta Q^2\rangle/[N\bar t]$,
where $\delta Q$ is the fluctuation of the charge, transmitted during the interval of time $N\bar t$, and the angular brackets denote the average over all the realizations of the noise. To find $\delta Q$ for a particular realization, we define $t_N=N\bar t+\Delta t$ as the time during which the charge $Ne$ is transferred. Then the charge, transmitted during the time $N\bar t$, equals $Q\approx Ne-I\Delta t$ with $I$ from Eq. (\ref{9}). Hence, $\delta Q=-I\Delta t$ and $S=2I^2\langle \Delta t^2\rangle/[N\bar t]$. Observe that $\Delta t\sim\sqrt{N}$. Thus, $S=2I^2\overline{\delta t^2}/\bar t$, where $\overline{\delta t^2}$ is the mean square fluctuation of the time required for 4 consecutive tunneling events. The latter fluctuation is computed in the same way as $\bar t$. One finds $S=2e^* I$ with the Fano factor

\begin{equation}
\label{10}
e^*=\frac{e\sum p_i}{64}\sum\frac{1}{p_i}=\frac{e}{8}\frac{2-s^2}{1-s^2+\frac{s^4}{8}[1-\cos(2\pi \Phi/\Phi_0+4\gamma)]},
\end{equation}
where $i$ runs over the set $\{ 0,\pi/2,\pi,-\pi/2\}$ and $s=2u|\Gamma_1\Gamma_2|/[|\Gamma_1|^2+|\Gamma_2|^2]$. At a low voltage bias, $s=1$ in a symmetric interferometer with $|\Gamma_1|=|\Gamma_2|$. Then the Fano factor
$e^*= e/[1-\cos(2\pi \Phi/\Phi_0+4\gamma)]$ diverges at $\Phi=[n-2\gamma/\pi]\Phi_0$.

The 113 topological order \cite{yf14} can also explain the observed upstream neutral mode and tunneling exponent. The explanation, based on the PH-Pfaffian state, has two advantages. First, the quantization of the Hall conductance in the 113 state depends on an edge equilibration mechanism. Any such mechanism fails at a sufficiently low temperature but no significant deviations from $G=5e^2/2h$ at low temperatures have been reported in the literature. At the same time, the conductance of the PH-Pfaffian state remains quantized arbitrarily close to absolute zero. Besides, an elegant combination of symmetry and non-Abelian statistics makes the PH-Pfaffian order aesthetically appealing.

In conclusion, the PH-Pfaffian order is consistent with all the  transport experiments at $\nu=5/2$. Numerical support for the Pfaffian and anti-Pfaffian states in simplified systems without disorder and LLM suggests an interesting possibility of ``symmetry from no symmetry". A smoking gun evidence of such an effect would come from a unique behavior in Mach-Zehnder interferometry.

We acknowledge support by the NSF under Grant No. DMR-1205715.


\begin{thebibliography}{99}
\bibitem{footnote-phase} The antiunitary particle-hole transformation changes a wave function of electrons into the complex conjugate wave function of holes. Hence, all statistical phase factors become complex conjugated.
\bibitem{footnote-edge} The decay length of unprotected neutral modes is determined by operators which are relevant in the renormalization group sense, in contrast to protected modes \cite{KF}.
\bibitem{footnote} The best estimates for $g$ are obtained by fitting the $I-V$ curve at a fixed quasiparticle charge.
\end{thebibliography}
\end{document}